# Field of View and contrast limitations of stellar interferometers. A quick review

François Hénault

*Institut de Planétologie et d'Astrophysique de Grenoble*
*Université Grenoble-Alpes, Centre National de la Recherche Scientifique*
*B.P. 53, 38041 Grenoble – France*

## ABSTRACT

Field of View (FoV) and contrast limitations of stellar interferometers have been the scope of numerous publications for more than thirty years. Recently, this topic regained some interest since long-baseline terrestrial interferometers or space borne nulling interferometers are envisioned for detecting and characterizing extra-solar planets orbiting in the habitable zone of their parent star. This goal supposes to achieving sufficient contrast ratio in the high angular frequency domain, thus on the whole interferometer FoV. In this paper are reviewed some of the contrast and FoV limiting factors, including spectral bandwidth, flux mismatches, fringe tracking, telescope image quality, atmosphere seeing, optical conjugation mismatch of the telescopes pupils, influence of anamorphous optics, pupil aberrations, signal-to-noise ratio and deviations with respect to "the golden rule of imaging interferometers". Finally, a tentative classification of all these factors is provided.

## 1 INTRODUCTION

Field of View (FoV) and contrast limitations of stellar interferometers have been the scope of numerous publications for more than thirty years. Recently, this topic regained some interest since long-baseline terrestrial interferometers or space borne nulling interferometers are envisioned for detecting and characterizing extra-solar planets orbiting in the habitable zone of their parent star. This goal supposes to achieving sufficient contrast ratio in the high angular frequency domain, thus on the whole interferometer FoV. In this paper are reviewed some of the contrast and FoV limiting factors. It starts with general mathematical framework described in section 2. In section 3 are evaluated the effects of various factors, including spectral bandwidth, flux mismatches, fringe tracking, telescope image quality, atmosphere seeing, optical conjugation mismatch of the telescopes entrance and exit pupils, influence of anamorphous optics, pupil aberrations and Signal-to-Noise Ratio (SNR), either by use of analytical expressions or numerical simulations. Two tentative contrast budgets are given in section 4, one for a classical "visibility imaging" stellar interferometer and the other for a very high contrast instrument. A brief conclusion is drawn in section 5. The use of more elaborated techniques such as calibration on reference stars, phase closures, high dispersion spectroscopy or Kernel nulling is beyond the scope of the paper.

## 2 GENERALITIES

### 2.1 Mathematical relations

The main coordinate frames are illustrated in Figure 1-A. It is assumed that both the collecting apertures in the entrance pupil plane OXY have identical diameters $D$ centred at points $P_n$ (n = 1,2), and that they are optically conjugated with their associated combining apertures centred at points P'$_n$ in the O'X'Y' exit pupil plane. Hence their diameters $D'$ also are identical.

Making use of a formalism extensively described in Refs. [1–3] and simplifying it for the case of a two-telescope interferometer, the intensity distributions generated by a point sky object, formed in the detection plane and retro-projected back onto the sky write as:

$$I(\mathbf{s_O},\mathbf{s}) = \text{PSF}(\mathbf{s}-\mathbf{s_O}) \times \{a_1^2 + a_2^2 + 2a_1a_2\cos[\varphi_1 - \varphi_2 + k\xi(\mathbf{s_O},\mathbf{s})]\}, \tag{1}$$

where $\xi(\mathbf{s_O},\mathbf{s})$ is an Optical Path Difference (OPD) term equal to:

$$\xi(\mathbf{s_O},\mathbf{s}) = \mathbf{s_O}\,\mathbf{P_2P_1} - \mathbf{s}\,\mathbf{P'_2P'_1}/m, \tag{2}$$

and the following scientific notations are used (bold characters denoting vectors):

| | |
|---|---|
| $a_n$ | The amplitude transmission factor of the n$^{th}$ interferometer arm ($1 \leq n \leq 2$) |
| $\varphi_n$ | A phase-shift introduced along the n$^{th}$ interferometer arm for OPD compensation and/or nulling purposes ($1 \leq n \leq 2$) |
| $k$ | The wavenumber $2\pi/\lambda$ of the electro-magnetic field assumed to be monochromatic, and $\lambda$ is its wavelength |
| PSF($\mathbf{s}$) | The PSF of an individual collecting telescope defined as $\text{PSF}(\mathbf{s}) = |\text{PSA}(\mathbf{s})|^2$, being projected back onto the sky |
| PSA($\mathbf{s}$) | The complex amplitude formed by an individual collecting telescope in the image plane and projected back onto the sky. For an unobstructed pupil of diameter $D$, it is the classical Airy amplitude function, equal to $2J_1(\rho)/\rho$, where $\rho = kD\|\mathbf{s}\|/2$ and $J_1$ is the type-J Bessel function at the first order |
| $\mathbf{s}$ | A unitary vector of direction cosines $\mathbf{s} \approx (u,v,1)$ directed at any point in the sky (corresponding to any point M" in the image plane), where the angular coordinates $u$ and $v$ are considered as first-order quantities |
| $\mathbf{s_O}$ | A unitary vector of direction cosines $\mathbf{s_O} \approx (u_O,v_O,1)$ pointed at a given sky object |
| $\mathbf{OP_n}$ | A vector of Cartesian coordinates $[x_n, y_n, z_n]$ defining the center P$_n$ of the n$^{th}$ sub-pupil in the entrance pupil plane ($1 \leq n \leq 2$) |
| $B$ | The entrance baseline of the interferometer |
| $\mathbf{O'P'_n}$ | A vector of Cartesian coordinates $[x'_n, y'_n, z'_n]$ defining the center P'$_n$ of the n$^{th}$ sub-pupil in the exit pupil plane ($1 \leq n \leq 2$) |
| $B'$ | The exit baseline of the interferometer |
| $m$ | The optical compression factor of the system, equal to $m = D'/D = F_C/F$ where $F$ and $F_C$ respectively are the focal lengths of the collecting telescopes and of the relay optics (see Figure 1-B). |

Eqs. 1–2 actually define ideal fringe patterns without contrast losses, showing a maximal contrast ratio $C_0(\mathbf{s_O},\mathbf{s})$ writing as:

$$C_0(\mathbf{s_O},\mathbf{s}) = \frac{2a_1a_2}{a_1^2 + a_2^2}, \tag{3}$$

to which all error sources listed in section 3 will be added separately in section 3.

### 2.2 To Have and Have Not Modulation

Eqs. 1–2 also evidence the role played by phase modulation. Ideally, the maximal contrast in Eq. 3 can be achieved in two different ways:

- Either modulating temporally the phase difference $\varphi_1(t)-\varphi_2(t)$ with delay lines, and interlacing scientific exposures with phase modulation exposures in order to compensating for non common OPDs by hardware or software, with the major drawback of reducing the observation time significantly,
- Or if the interferometer incorporates a multi-axial combiner, transferring the spatial modulation information of the interference pattern into the Fourier space, and controlling the OPDs by spectral analysis [4], not requiring additional phase modulation sequences.

In both cases, significant improvements of the FoV and contrast factor may be achieved, either by hardware or software means. However such techniques may be affected with their own instrumental errors, and will not be discussed in the present paper.

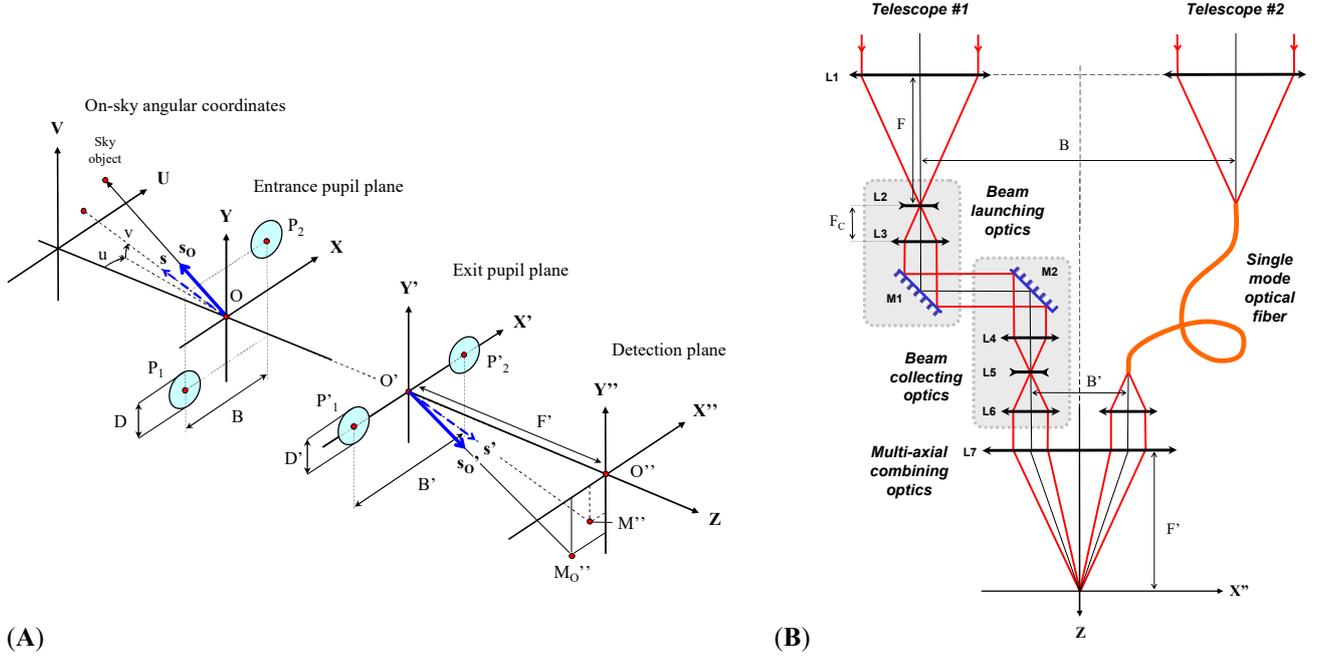

Figure 1: (**A**) Reference frames used on-sky (O,U,V), on the entrance pupil plane (O,X,Y) and on the exit pupil plane (O',X',Y'). The coordinate system (O'',X'',Y'') attached to the detection plane is optically conjugated with the (O,U,V) reference frame. (**B**) Schematic optical layout of a two-aperture interferometer, either including bulk or fiber optics.

## 3   CONTRAST LOSSES

Referring to stellar interferometry literature [5], there actually exists a large variety of contrast and/or FoV degradation factors, of which a tentative analysis is given in this section. Two of them, however, will not be discussed extensively here. They are:
- The FoV can only be limited by a field stop located at the image plane of the interferometer. This is an obvious limitation, but all existing interferometers include such components or their equivalent (single-mode fibers, V-groove arrays, etc.),
- Other interesting definitions may depend on the observed sky scene, such as defined in Ref. [6] as the Clean FoV (CLF) or Direct Imaging FoV (DIF). However the present study is restricted to point sky objects, either located on or off the axis of the interferometer.
- Finally, it must be noted that the FoV formulas given in Refs. [2–3] are correct mathematically, but did not cover any of the contrast and FoV limitations presented in the following sub-sections.

## 3.1 Spectral bandwidth

Eq. 1 shows that the interferometer fringe pattern $I(\mathbf{s_O},\mathbf{s})$ is fundamentally chromatic since it depends on the wavenumber $k = 2\pi/\lambda$. Thus a contrast loss occurs as discussed in Refs. [7–9]. Rigorously, all monochromatic interferograms in Eq. 1 should be integrated over a finite spectral bandwidth $\Delta\lambda$ that is $[\lambda–\Delta\lambda/2, \lambda+\Delta\lambda/2]$:

$$I_{\Delta\lambda}(\mathbf{s_O},\mathbf{s}) = \int_{\lambda-\Delta\lambda/2}^{\lambda+\Delta\lambda/2} I(\mathbf{s_O},\mathbf{s})\, d\lambda. \tag{4}$$

Assuming that $\Delta\lambda$ is sensibly smaller than $\lambda$, a Taylor development at the first-order leads to an approximate expression of $I_{\Delta\lambda}(\mathbf{s_O},\mathbf{s})$:

$$I_{\Delta\lambda}(\mathbf{s_O},\mathbf{s}) \approx I(\mathbf{s_O},\mathbf{s}) \operatorname{sinc}\left[\pi\, \Delta\lambda\, \xi(\mathbf{s_O},\mathbf{s})/\lambda^2\right], \tag{5}$$

with sinc($u$) being the sine cardinal function equal to sin($u$)/$u$. Thus the contrast factor is equal to $C(\mathbf{s_O},\mathbf{s}) = \operatorname{sinc}\left[\pi\, \Delta\lambda\, \xi(\mathbf{s_O},\mathbf{s})/\lambda^2\right]$ and stands for an intrinsic limitation of the experiment in view of its scientific goals. It should be noted however that the fringe pattern itself can be made achromatic by inserting dispersive components into intermediate image planes, leading to a FANI solution that was described in Ref. [9] and is illustrated in Figure 2 for the case of a nulling Bracewell interferometer.

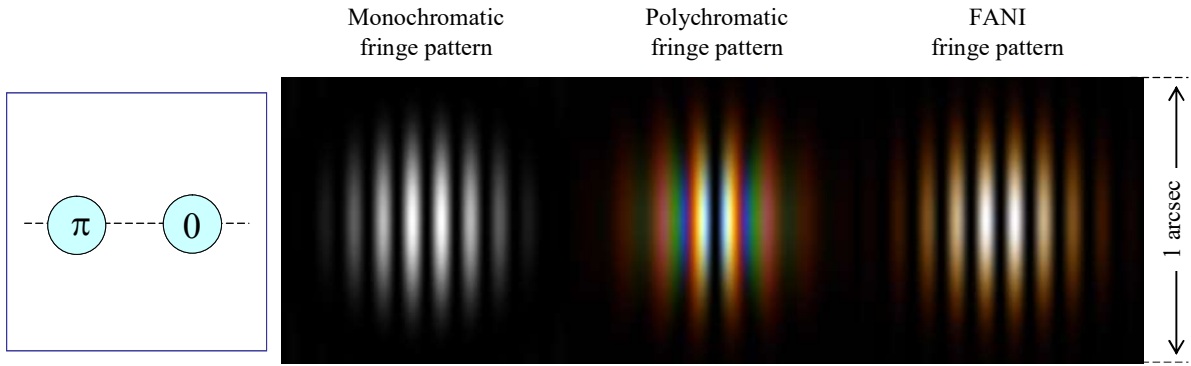

Figure 2: Illustrating achromatized fringe patterns achieved with the FANI optical design [9]. A pseudo-colour scale indicates which kind of photons is predominant, blue for lowest wavelengths and red for highest ones.

## 3.2 Flux mismatch

Setting $a_1 = a$ and $a_2 = a + da$ into Eq. 3 readily leads to a simple expression of the contrast loss due to flux mismatch:

$$C(\mathbf{s_O},\mathbf{s}) = 1 - (da)^2 = 1 - dI, \tag{6}$$

where $dI = |I_1 - I_2|$ is the flux difference between the sub-pupils. However, it must be emphasized that evaluating this contrast loss may not be as simple as it seems, because it should include polarization effects occurring from high incidence angles on the mirrors, especially those located into the relay optics subsystems. A detailed discussion about polarization effects in interferometry is out of the scope of this paper, but can be found in Ref. [10]. Here is simply assumed that they are included into the global parameter $dI$.

> From now and in order to alleviating the analytical formulas, it is assumed that $a_1 = a_2 = 0.5$ in the remainder of the text.

### 3.3 Fringe tracking

Assuming that fringe tracking is carried out on a star located at the FoV centre, $s_O$ is equal to the null vector and Eq. 1 reduces to:

$$I(s) = PSF(s) \frac{1 + \cos[\varphi_1(t) - \varphi_2(t) + k\, s\, P'_2\, P'_1 / m]}{2}, \quad (7)$$

and in the absence of any temporal or spatial modulation (see § 2.2) the contrast factor writes as:

$$C(s) = \cos[\varphi_1(t) - \varphi_2(t)] = \cos[k\delta_{12}(t)], \quad (8)$$

with $\delta_{12}(t)$ the differential OPD between both interferometer arms.

### 3.4 Telescope image quality

The Wavefront errors (WFE) of each individual telescope and of its downstream optical train are an important source of contrast degradation and FoV limitation. They can be evaluated by generalizing Eq. 2 as:

$$\xi(s_O, s) = s_O\, P_2\, P_1 - s\, P'_2\, P'_1 / m + s\, Q'_2\, Q'_1, \quad (9)$$

where Q'$_1$ and Q'$_2$ are points located inside the exit sub-pupils n°1 and 2 as illustrated on the left side of Figure 3. Denoting their Cartesian coordinates [d$x'_n$, d$y'_n$, $\Delta_n(Q'_n)$] with $\Delta_n(Q'_n)$ the WFE across the n$^{th}$ exit sub-pupil ($1 \leq n \leq 2$), the complex amplitude distribution formed in the image plane writes as:

$$PSA_n(s) = \iint_{Q'_n} B_{D'}(Q'_n) \exp[ik(u\, dx'_n + v\, dy'_n)] \exp[ik\, \Delta_n(Q'_n)] dQ'_n, \quad (10)$$

where $B_{D'}(Q'_n)$ is the "pillbox" function equal to unity if $\|P'_n Q'_n\| \leq D'/2$ and to zero elsewhere. Now summing $PSA_1(s)$ and $PSA_2(s)$ coherently and taking the square modulus of the result leads to the following expression of the fringe pattern:

$$I(s, s_O) = \frac{|PSA_1(s)|^2 + |PSA_2(s)|^2 + 2\, PSA_1(s) \times cc[PSA_2(s)] \cos[\varphi_1 - \varphi_2 + k\xi(s_O, s)]}{2}, \quad (11)$$

where $cc[z]$ denotes the complex conjugate of $z$. Eq. 11 has no simple analytical expression, however assuming weak WFE aberrations and developing $\exp[ik\, \Delta_n(Q'_n)] \approx 1 + ik\, \Delta_n(Q'_n)$ allows approximating $PSA_n(s)$ as:

$$PSA_n(s) \approx \hat{B}_{D'}(s) + ik\hat{\Delta}_n(s), \quad (12)$$

where $\hat{B}_{D'}(s)$ and $\hat{\Delta}_n(s)$ are the Fourier transforms of $B_{D'}(Q'_n)$ and $\Delta_n(Q'_n)$ respectively. Then the contrast factor of the interference pattern can be written as:

$$C(s_O, s) \approx \frac{\sqrt{PSF^2(s) + k^2 PSF(s) \times [\hat{\Delta}_1^2(s) + \hat{\Delta}_2^2(s)] + k^4 \hat{\Delta}_1^2(s) \hat{\Delta}_2^2(s)}}{PSF(s) + k^2 \hat{\Delta}_1^2(s)/2 + k^2 \hat{\Delta}_2^2(s)/2}, \quad (13)$$

with the noticeable drawback that the dependence of $C(s_O, s)$ on vector $s_O$ disappears. However, one remarkable property of Eq. 13 is that the contrast ratio reduces to unity when $\hat{\Delta}_1(s) = \hat{\Delta}_2(s)$ and consequently $\Delta_1(Q'_n) = \Delta_2(Q'_n)$, i.e. the WFE aberrations are identical along both interferometer arms. Further simplification of Eq. 13 occurs with the reasonable assumption that the $\hat{\Delta}_n(\mathbf{s})$ functions are nearly constant and equal to $\hat{\Delta}_n(\mathbf{0})$ over the whole interferometer FoV, finally leading to:

$$C(\mathbf{s}) \approx \mathrm{P}S\mathrm{F}(\mathbf{s}) \times \left(1 - k^2 \overline{\Delta}_1^2/4 - k^2 \overline{\Delta}_2^2/4\right), \quad \text{with:}$$
$$\overline{\Delta}_n = \iint_{Q'_n} B_{D'}(Q'_n)\Delta_n(Q'_n)dQ'_n \bigg/ \iint_{Q'_n} B_{D'}(Q'_n)dQ'_n \quad \text{and when:} \quad \Delta_1(Q'_n) \neq \Delta_2(Q'_n). \quad (14)$$

In Figure 3 is illustrated the contrast degradation due to telescope image quality in the case of a defocus WFE varying from 0 to $0.5\lambda$ RMS, either when the WFEs are identical on both interferometer arms (top of the Figure) or different (bottom). It can be seen that the contrast factor is only governed by the shape of the fringes envelope $\mathrm{P}S\mathrm{F}(\mathbf{s})$ in the first case, while drastic and immediate contrast losses and FoV reduction occur in the second case. It must be noted that these interference patterns were generated using direct numerical simulations of Eq. 11, and not the following approximate relations 12–14. The employed numerical parameters are summarized in Table 1

Table 1: Used parameters for numerical simulations.

| Parameter | Value |
|---|---|
| Wavelength | $\lambda = 10$ μm |
| Diameter of entrance sub-pupils | $D = 1$ m |
| Baseline of entrance sub-pupils | $B = 100$ m |
| Telescopes focal length | $F = 50$ m |
| Relay optics focal length | $F_C = 0.1$ m |
| Combiner focal length | $F' = 1$ m |
| Diameter of exit sub-pupils | $D' = 1$ mm |
| Baseline of exit sub-pupils | $B' = 5$ mm |

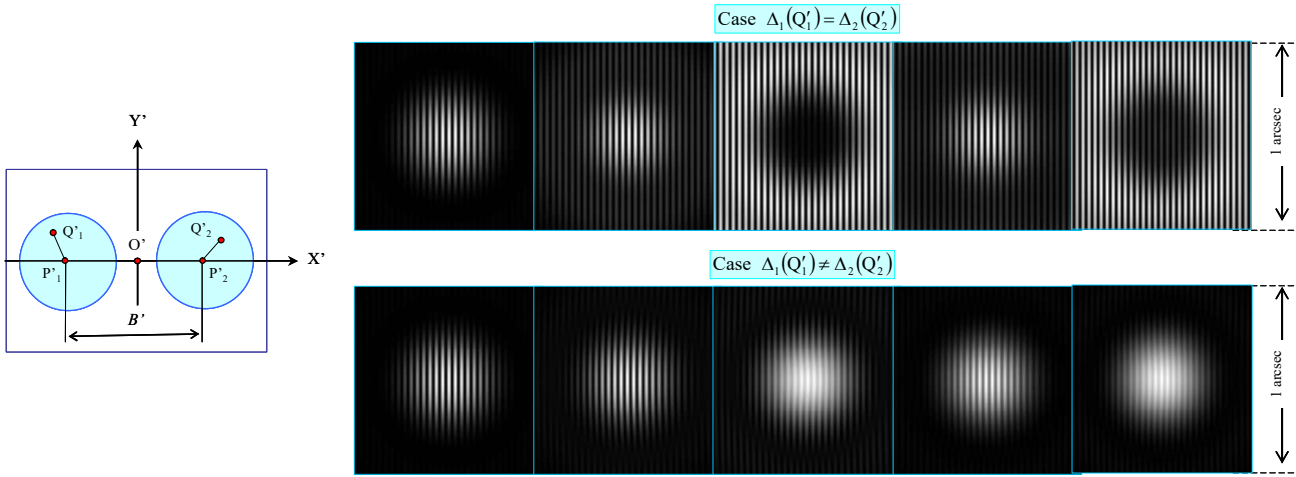

Figure 3: Illustrating the contrast degradation due to telescope image quality. Left side: Exit pupil geometry. Right side, top: Effect of a defocus WFE varying from 0 to $0.5\lambda$ RMS, case of identical WFEs on both interferometer arms. Right side, bottom: Same case with non common WFEs.

### 3.5 Atmosphere seeing

Atmosphere seeing is perhaps the major cause of contrast degradation in stellar interferometry. Basically, this case may be handled in the same way as in the previous sub-section. Here the WFE aberrations are so strong that the approximate relations 13–14 do not hold, but they remain applicable to Adaptive Optics (AO) corrected systems [11]. The parameters $\overline{\Delta}_n$ then stand for residual WFEs after AO correction. Figure 4 illustrates the intensity distributions in the image plane for residual AO errors equal to $0.25\lambda$ RMS. Here again contrast losses and FoV reduction appear rapidly in the general case when $\Delta_1(Q'_n) \neq \Delta_2(Q'_n)$. The employed numerical parameters are those summarized in Table 1.

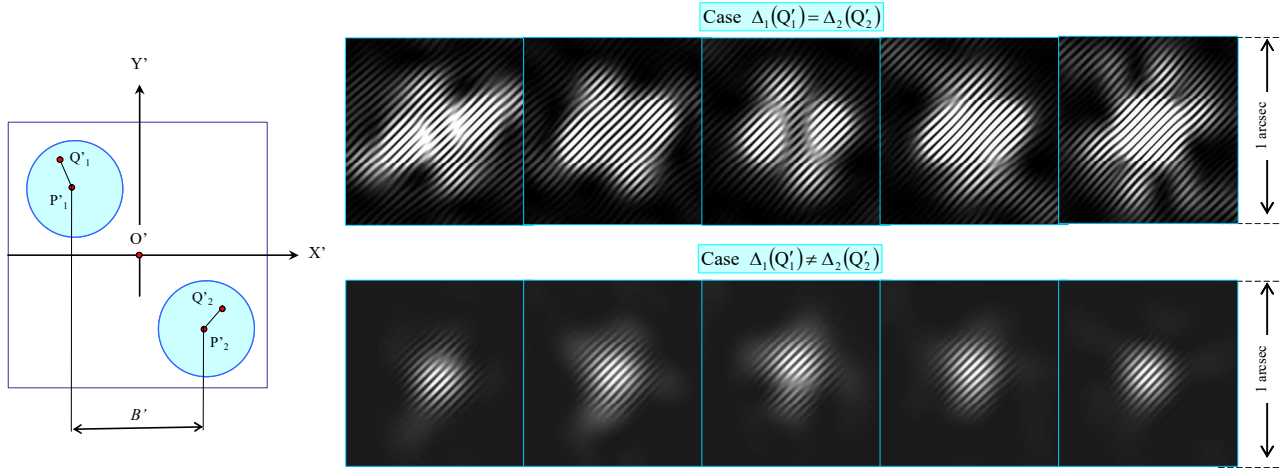

Figure 4: Same illustrations as in Figure 3 with an atmosphere seeing of $0.25\lambda$ RMS.

### 3.6 Pupils conjugation mismatch

The effect of pupils conjugation mismatch has been extensively discussed in Refs. [3] and [12]. They may originate from two different causes as illustrated in Figure 5.
- For a ground stellar interferometer the entrance sub-pupils are usually located at the collecting telescopes pupils. When staring at an off-axis target, they do not lie in the same plane and wander around their mean position O by quantities noted $z_1$ and $z_2$ (see Figure 5-a).
- Additional pupil location mismatches may be introduced by the relay optics between the collecting telescopes and the combiner, noted $z'_1$ and $z'_2$ in Figure 5-b.

Their main effect is to generate curved fringe patterns such as illustrated in Figure 5-c. In that case Eqs. 1–3 are still applicable with unitary vectors $\mathbf{s_O}$ and $\mathbf{s}$ developed at the second order, i.e. $\mathbf{s_O} \approx [u_O, v_O, 1 - u_O^2/2 - v_O^2/2]$ and $\mathbf{s} \approx [u, v, 1 - u^2/2 - v/2]$, thus the contrast ratio remains equal to unity. $\Delta_1(Q'_n) \neq \Delta_2(Q'_n)$. The parameters employed for numerical simulations are those summarized in Table 1.

### 3.7 Anamorphous optics

Most of intererometric combiners incorporate anamorphous optics stretching the fringe pattern along one axis (noted Y" in Figure 1) in order to minimizing the number of required pixels on the detector array, especially for spectroscopy applications. In that case the analytical expressions in Eqs. 1–2 must be integrated and averaged over that axis. The mathematical development is not really difficult, but very long and cumbersome. After matching PSF(**s**) to a Gaussian distribution and expanding the result to the second order with respect to $u_0$, it finally leads to an approximate expression of the contrast factor as:

$$C(\mathbf{s_o},\mathbf{s}) \approx 1 - 2u_O^2 \frac{2.25 - k^2 dz_{21}'^2/8m^2}{1 + 0.188\,kR - 22.2/k^2R^2} \tag{15}$$

with $R$ the radius of an individual entrance sub-pupil and $dz_{21}' = z_2' - z_1'$. The anamorphosed fringe patterns shown in Figure 5-c are depicted in Figure 5-d. The employed numerical parameters are those summarized in Table 1.

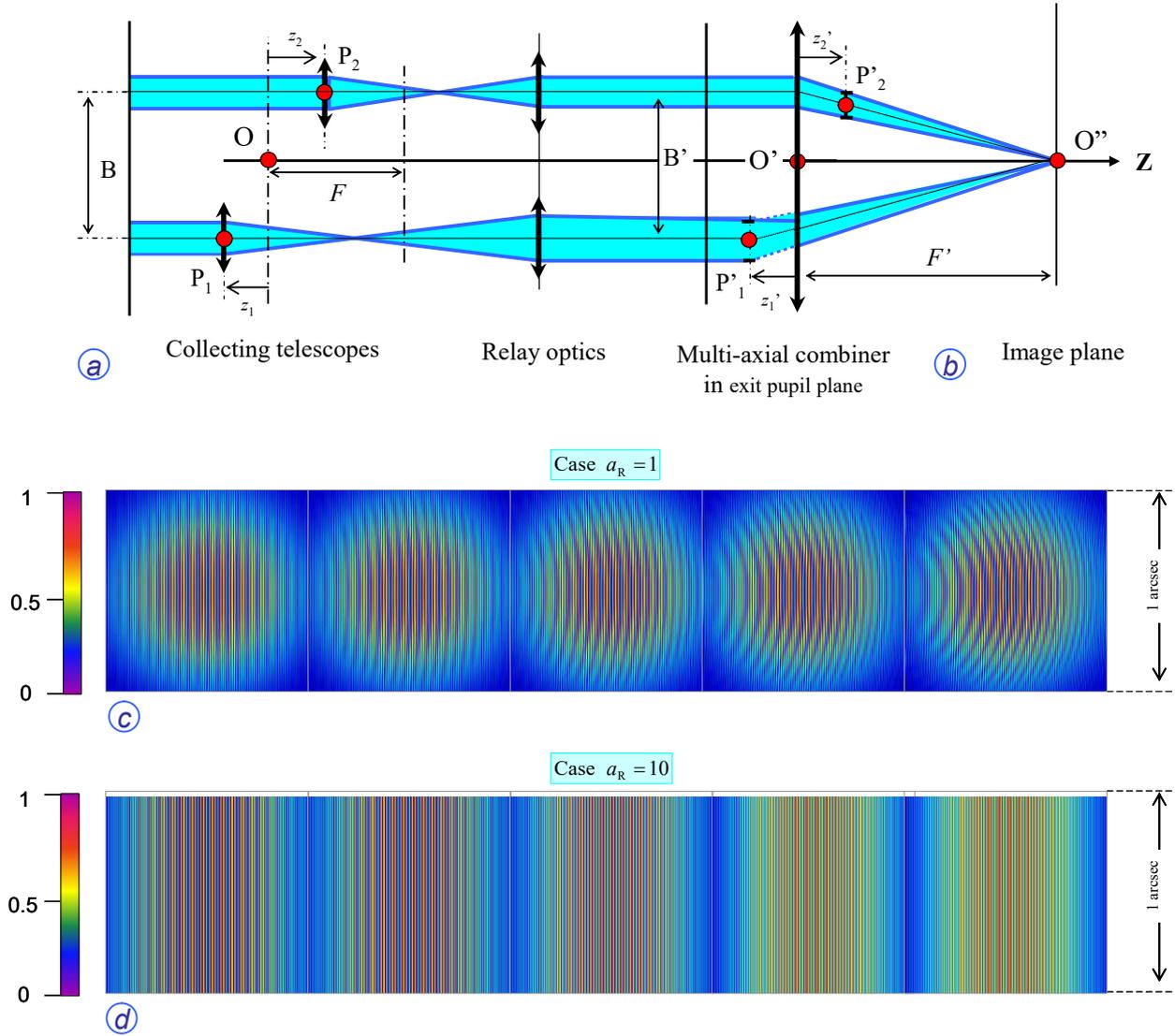

Figure 5: Illustrating the contrast degradation due to conjugation mismatch of the telescope pupils. (*a*) With respect to entrance pupil plane. (*b*) With respect to exit pupil plane. (*c*) Raw intensity distributions of the fringe patterns. (*d*) Same patterns as in (*c*) anamorphosed by a factor $a_R = 10$ along the vertical axis. Fringe patterns have been re-squared for the sake of visual illustration.

### 3.8 Pupil aberration

While both previous subsections dealt with pupil conjugation mismatches that are well documented, it seems that so far little attention was paid to their own pupilar aberrations. Those aberrations are naturally generated by the relay optics from the collecting telescopes to the final combiner, and their main effect will be to blurring the exit sub-pupils of the interferometer, possibly in a different way as illustrated in Figure 6.

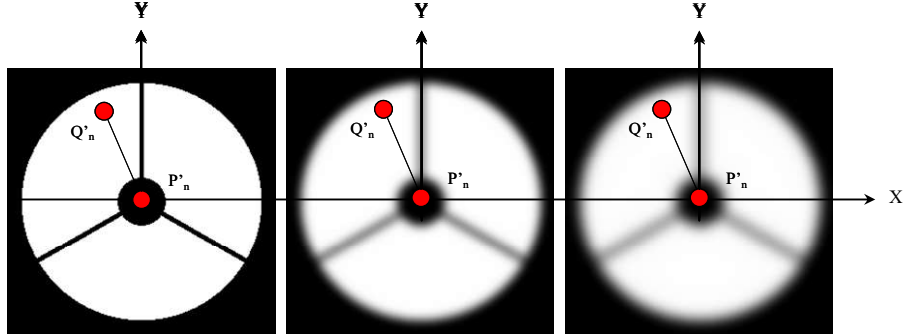

Figure 6: Illustrating pupil aberration blur effects.

Let us assume that the blur effect is modelled by a Gaussian function $G'(Q'_n)$ of standard deviation σ in the exit pupil plane O'X'Y'. It writes as:

$$G'(Q'_n) = \frac{\exp(-Q'^2_n/2\sigma^2)}{\iint_{Q'_n} \exp(-Q'^2_n/2\sigma^2) dQ'_n} \quad \text{with} \quad (1 \leq n \leq 2). \quad (16)$$

Here again the resulting contrast ration does not depend on vector $\mathbf{s_O}$ and can be expressed as:

$$C(\mathbf{s}) = \exp\left[-2\sigma^2 \pi^2 \|\mathbf{s}\|^2 / \lambda^2\right] = \exp\left[-k^2 \sigma^2 \|\mathbf{s}\|^2 / 2\right]. \quad (17)$$

### 3.9 Signal-to-Noise Ratio (SNR)

Building a detailed SNR budget of the interferometer including shot noise, read-out noise, dark current, digitization noise etc. is beyond the scope of this paper. Instead, the SNR is considered as a macroscopic parameter gathering all noises together. Adding white noise to the intensities $I_1$ and $I_1$ recorded along both interferometer arms (see § 3.2) and developing Eq. 1 at the second order, the contrast ratio can be approximated in a simple form:

$$C(\mathbf{s}) \approx 1 - \frac{2}{SNR^2} \quad (18)$$

### 3.10 "Golden rule" of imaging interferometers

A "golden rule" for imaging interferometers states that their exit sub-pupils should be a homothetic replica of their entrance sub-pupils in order to maximizing the FoV and contrast [13]. It has been the subject of extensive literature (see Ref. [1] and the cited references therein) and its applicability to non-imaging Michelson interferometers is discussed in Ref. [14]. Practically, most of the existing interferometers do not respect this golden rule, at the noticeable exception of the Large Binocular Telescope Interferometer (LBTI).

Figure 7 illustrates the effect of such a violation, inspired from the VEGA/CHARA interferometer combiner [4]. The employed numerical parameters are those summarized in Table 1.

In Figure 7 are displayed the geometric configurations of the entrance and exit sub-pupils of the interferometer, showing an evident violation of the golden rule. Figure 7-c to 7-e depict the intensity distributions of the fringe patterns formed by three different sky objects, either located on or off-axis of the interferometer. The numerical simulations evidence a slight contrast loss induced by the anamorphosing optics, about 2-4% depending on the location of the sky objects.

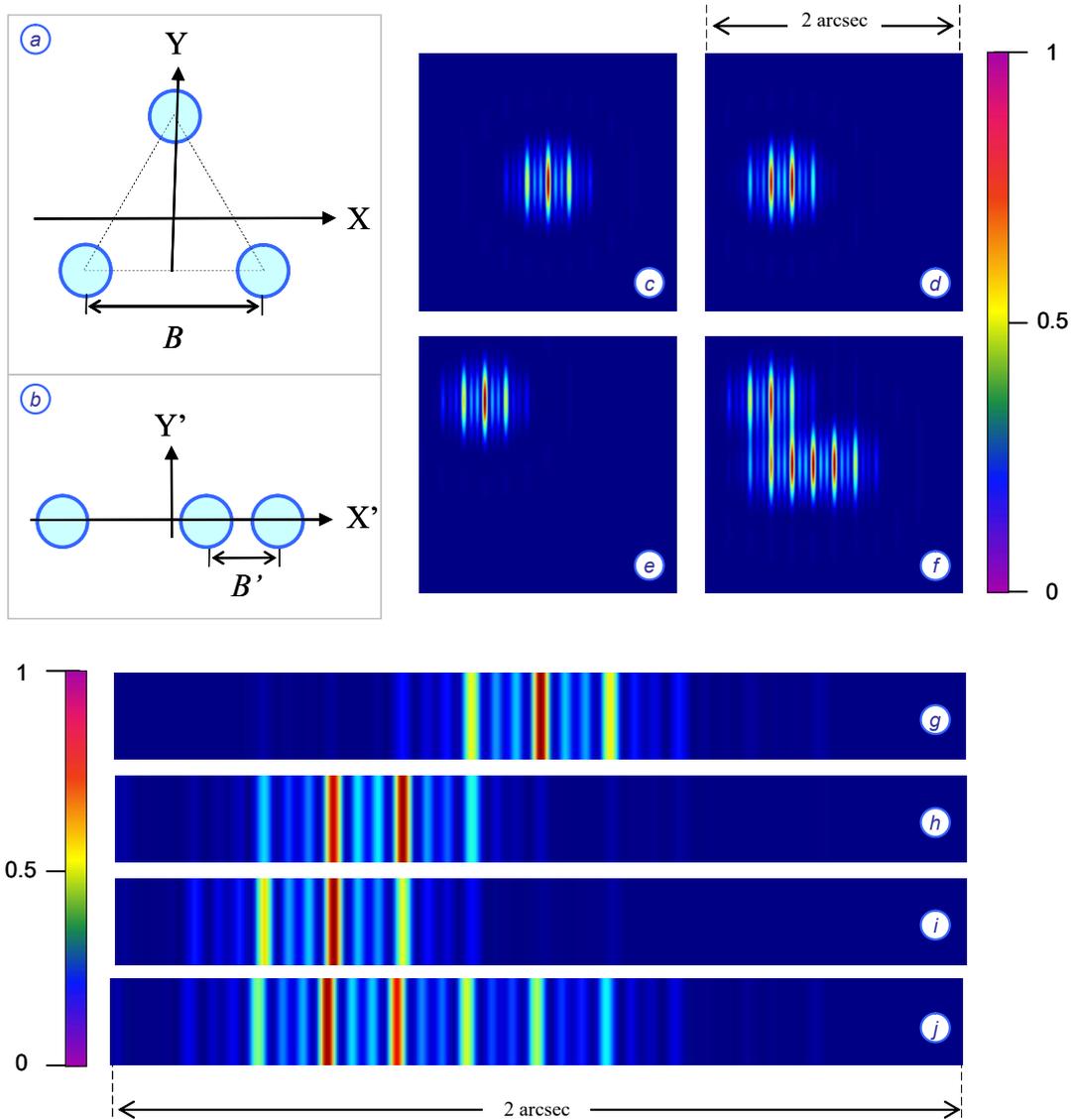

Figure 7: Illustrating the effect of a golden rule violation. (*a*) Geometry of the entrance pupils. (*b*) Geometry of the exit pupils. (*c*) Fringe pattern generated by an on-axis sky object. (*d*) Fringe pattern generated by an off-axis sky object located at angular coordinates [–0.8, 0] arcsec. (*e*) Fringe pattern generated by an off-axis sky object located at angular coordinates [–1, +1] arcsec. (*f*) Simultaneous observation of the three previous sky objects. (*g–j*) Same fringe patterns as (*c–f*) anamorphosed by a factor $a_R = 10$ along the vertical axis.

# 4 TENTATIVE BUDGETS

In Table 2 are presented two tentative contrast budgets for both the cases of a medium and a high contrast stellar interferometer. The "medium" one corresponds to usual visibility measurements, where it is assumed that a global contrast goal of 0.8 is sufficient. The "high contrast" case is much more demanding, since it aims at detecting and characterizing an extra-solar planet $10^{-4}$ fainter than its parent star. Hence a minimal contrast ratio higher than 0.999 is required. The following hypotheses are made:

- Only the case of a two-telescope interferometer is considered, whose parameters are summarized in Table 1,
- Contrast factors are purely monochromatic, thus polychromatic contrast losses defined by Eq. 4 should be added depending on the science applications,
- Most of the contributors are evaluated using the analytical relations given in subsections 3.2 to 3.9, excepting the image quality and pupil location items that are computed numerically,
- Most of them are computed for a sky object located on-axis, i.e. $\mathbf{s_O} = \mathbf{0}$, excepting pupil locations mismatches and pupil aberrations where the sky object is placed at 1 arcsec off-axis;
- No deviation with respect to the golden rule of imaging interferometers is assumed, thus the contrast losses estimated in § 3.10 are neglected.
- The telescope image quality and AO residual WFEs are separated in order to taking into account Non common path aberrations (NCPA) of the system,
- Anamorphous optics are present near the image plane with a ratio $a_R = 10$,
- Finally, no dynamic temporal or spatial modulation processes are applied.

The most critical contributors are found to be the WFEs resulting from telescope aberrations, AO residuals or NCPAs, and the pupil locations mismatches. Not surprisingly, these system requirements are much tougher for the extra-solar planet detection case, where the flux mismatches and pupil aberrations must not be neglected. It must be noted that the requirements could be somewhat relaxed in the absence of anamorphous optics, in which case the entrance and exit pupil location mismatches may be set to unity.

Table 2: Tentative contrast budgets for both the cases of medium and high contrast stellar interferometers.

|  | | Parameters | | | Contrast | |
| --- | --- | --- | --- | --- | --- | --- |
|  | Name | Medium contrast | High contrast | Unit | Medium contrast | High contrast |
| Flux mismatch | $dI$ | 10 | 0.01 | % (*) | 0.90 | 0.9999 |
| Fringe tracking | $\delta_{12}(t)$ | 200 | 20 | nm | 0.99 | 0.9999 |
| Telescope image quality | $\Delta_n$ | 0.25 | 0.025 | $\lambda$ RMS | 0.99 | 0.9999 |
| AO residual WFE | $\Delta_n$ | 0.25 | 0.025 | $\lambda$ RMS | 0.99 | 0.9999 |
| Entrance pupil locations | $dz_{21}$ | 40 | 1 | m | 0.97 | 1.0000 |
| Exit pupil locations | $dz'_{21}$ | 0.025 | 0.002 | m | 0.99 | 0.9999 |
| Pupil aberration | $\sigma$ | 10 | 1 | % (**) | 0.97 | 0.9997 |
| Signal-to-Noise Ratio | $SNR$ | 10 | 100 | - | 0.98 | 0.9998 |
| (*) with respect to flux impinging on a single telescope | | | | **Global** | **0.80** | **0.9990** |
| (**) with respect to entrance pupil diameter | | | | **Goal** | **0.80** | **0.9990** |

## 5 CONCLUSION

In this paper were reviewed the main FoV and contrast limitations of a stellar interferometer, which were the scope of numerous publications for more than thirty years [1–14]. Nowadays this topic regains interest since long-baseline terrestrial interferometers or space borne nulling interferometers are envisioned for detecting and characterizing extra-solar planets orbiting around their parent star. This very ambitious goal requires to achieving high contrast rates on the whole interferometer FoV. Some of the most critical contrast and FoV limiting factors were summarized in section 3, including spectral bandwidth, flux mismatches, fringe tracking, telescope image quality, atmosphere seeing, optical conjugation mismatch of the telescope pupils, influence of anamorphous optics, pupil aberrations, signal-to-noise ratio, and deviations with respect to the golden rule of imaging interferometers. Two tentative contrast budgets were presented in section 4, one for a classical visibility imaging stellar interferometer and the other for a very high contrast instrument. The most critical contributors are found to be to the WFEs resulting from telescope aberrations, AO residuals and NCPAs. Not surprisingly, the system requirements are much tougher for the extra-solar planet detection case, where flux mismatches and pupil aberrations cannot be neglected. It must be noted that the presence of anamorphous optics near the focal plane of the interferometer worsens at least two contrast loss factors related to both the entrance and exit pupils location mismatch (§ 3.7) and violation of the golden rule (§ 3.10). This pleads in favour of replacing the anamorphous optics with an Integral field unit (IFU), with the drawback of more complex design and more voluminous combiner, of the class of the MUSE instrument installed at the VLT [15–16]. It might be the price to pay for achieving very high contrast interferometric measurements.